\documentclass[12pt,preprint]{aastex}
\usepackage{float}

\shorttitle{Early Swift Afterglows}
\shortauthors{Butler \& Kocevski}

\def\gtrsim{\mathrel{\hbox{\rlap{\hbox{\lower4pt\hbox{$\sim$}}}\hbox{$>$}}}}
\def\lessim{\mathrel{\hbox{\rlap{\hbox{\lower4pt\hbox{$\sim$}}}\hbox{$<$}}}}
\newcommand{\beq}{\begin{equation}}
\newcommand{\eeq}{\end{equation}}

\newcommand\swift{{\it Swift}}

\slugcomment{Submitted to ApJ}

\begin{document}

\title{X-ray Hardness Evolution in GRB Afterglows and Flares: \\ Late Time GRB
Activity Without $N_H$ Variations}
\author{Nathaniel R. Butler\altaffilmark{1,2} and
Daniel Kocevski\altaffilmark{2}}
\altaffiltext{1}{Townes Fellow, Space Sciences Laboratory,
University of California, Berkeley, CA, 94720-7450, USA}
\altaffiltext{2}{Astronomy Department, University of California,
445 Campbell Hall, Berkeley, CA 94720-3411, USA}

\begin{abstract}
We show that the X-ray and $\gamma$-ray spectra of Swift GRBs and their
afterglows are consistent with the emission characteristic of an expanding, relativistic
fireball.  The classical afterglow due to the impact of the fireball on the external
medium is often not observed until one to several hours after the GRB.
Focusing on GRBs 061121, 060614, and 060124, but generalizing to the full ($>$50 Msec
XRT exposure) Swift sample up to and including GRB~061210,
we show that the early emission in $>$90\% of early afterglows
has a characteristic $\nu F_{\nu}$ spectral energy $E_{\rm peak}$
which likely evolves from the $\gamma$-rays through the soft X-ray band on timescales of $10^2-10^4$s 
after the GRB.  The observed spectra are strongly curved when plotted with logarithmic axes and have often been incorrectly
fitted in other studies with a time-varying soft X-ray absorption.  The spectral 
evolution inferred from fitting instead models used to fit GRBs demonstrates a common evolution---a
powerlaw hardness intensity correlation and hard to soft evolution---for GRBs and the early X-ray
afterglows and X-ray flares.  Combined with studies of short timescale variability, our
findings indicate a central engine active for longer than previously suspected.  The GRB spectra
are observed to become very
soft at late times due to an intrinsic spectral evolution and due to the surprising faintness
of some afterglows.  We discuss models for the early X-ray emission.
\end{abstract}

\keywords{gamma rays: bursts --- supernovae: general --- X-rays: general}

\maketitle

\section{Introduction}
\label{sec:intro}

The Swift satellite \citep{gehrels04} and its X-ray telescope \citep{burrows05b} have opened a new window
into the early lives of $\gamma$-ray Bursts (GRBs) and their afterglows.  We see
a complex array of behaviors, many of which appear to directly conflict \citep[e.g.,][]{obrien06,pain06,willin06} the well
tested internal--external shock GRB and afterglow model \citep{rnm94,snp97,spn98,wng99}.  In this ``fireball''
model, the GRB is produced via collisions of shells in a relativistic outflow, and an afterglow
arises later as the ejecta sweep up and heat the surrounding medium.  The Swift afterglows exhibit dramatic
flaring, rapidly decaying prompt emission tails, and typically a broad plateau phase until $t\approx 10^4$s \citep[e.g.][]{nousek06}.
Early afterglow observations prior to Swift \citep[e.g.,][]{frontera2000} suggested instead
a $\sim 10$s duration burst rapidly gone and replaced by the fading afterglow emission.  How these
observations are to be reconciled and what mechanisms produce
the early afterglow emission are key open questions.

Particularly intriguing,
several recent studies fit the {\it Swift}~X-ray Telescope (XRT) data and infer a time variable soft X-ray
absorption \citep{starling05,rol06,campana06c}. 
This would imply that the early afterglow is stripping electrons from a dense shell of light-element-rich material
located $R\lessim 1$ pc from the GRB, which was not already fully ionized by the GRB.  It is difficult to detect
such an effect because of the strong spectral evolution common in the early afterglows
\citep[e.g., ][``Paper I'']{vaughan06,butler07a}.  A changing column density $N_H$ cannot easily be separated from intrinsic 
afterglow spectral evolution, given the narrow XRT bandpass.  If the early X-ray spectra exhibit log-log curvature like that of 
GRBs, which have $\nu F_{\nu}$ spectral turnovers at a characteristic energy $E_{\rm peak}$ \citep[e.g.,][]{preece00,kaneko06}, then
evolution in the curvature could be mistaken for variations in $N_H$.

As we discuss below, plots of early XRT spectra do show strong log-log curvature and an inferred $E_{\rm peak}$ which typically passes in time
through the X-ray band.
This produces a changing X-ray hardness, which we observe to correlate with the flux.
A close analogy can be found in the spectral evolution of GRBs observed by the Burst and Transient Source Experiment \citep[BATSE;][]{fishman99}.
A characteristic feature of these spectra and light curves
is a hard-to-soft evolution in time and a powerlaw hardness--intensity correlation
\citep{gol83,karg95,norris96,fenimore95,fenimore96}.
The recent refined study of \citet{bnr01} measures a powerlaw relation between the characteristic energy
$E_{\rm peak}$ and the bolometric flux $F_{\rm bol}$ valid for $>$57\% of GRB pulses,
$E_{\rm peak} \propto F_{\rm bol}^{0.5\pm0.2}$.  We observe a consistent correlation in the soft, early-time XRT data.

In Paper I, we present evidence for this outlier population
of extremely soft afterglows in the first year of \swift~XRT
afterglow data.  Although they were
identified via an automated search for spectral lines,  
the spectra are also fitted well by models
containing multiple continuum components.  Below and in \citet{butNkoc07}, we explore further
the phenomenology associated with this soft emission.  We demonstrate that
GRB-like behavior is present in the first $t\lessim 1$ hour of
$>$90\% of the afterglows and is especially prominent during the flaring.
In two cases, thanks to Burst Alert Telescope (BAT) triggers on bright precursors,
X-ray emission coincident in time with the classical GRB is detected and can be shown
to have quite similar properties to the highly time-variable emission at later times.
This is strong evidence---to be combined with the short timescale variabiliity
studies \citep[e.g.,][]{burrows05a,falcone06,romano06a,pagani06,koc07}---tying
the flare and early afterglow emission to the GRB central engine.

\section{Data Reduction}

Our automated pipeline at U.~C. Berkeley downloads the \swift~data in near real
time from the {\it Swift}~Archive\footnote{ftp://legacy.gsfc.nasa.gov/swift/data}
and quicklook site.  
We use the calibration files from the 2006-04-27 BAT and XRT database 
release.
The additional automated processing described below is done uniformly for all afterglows
via custom IDL scripts.  The final data products are available for general
consumption\footnote{http://astro.berkeley.edu/$\sim$nat/swift}.

The XRT suffers from a significant number of bad or unstable pixels and columns.
Two central detector columns were lost due to a micro-meteorite 
strike\footnote{http://swift.gsfc.nasa.gov/docs/heasarc/caldb/swift/docs/xrt/SWIFT-XRT-CALDB-01\_v5.pdf}.  For the
early afterglows ($t\lessim 10^3$s), when the satellite initially points the XRT at the source without the precise localization
information needed to offset from the bad columns, a large and time-dependent fraction of the
flux can be lost.
In order to produce accurate light curves and properly normalized spectra, it
is necessary to accurately determine the position centroid and to precisely track the
loss of source and background flux due to the bad detector elements on short ($\sim$ few second)
timescales.

\subsection{Photon Counting (PC) Mode Light Curves}

  We begin by projecting the data in the 0.5-8.0 keV band from each PC mode followup observation onto a tangent plane
  centered at the source position quoted by the XRT Team.
  In raw coordinates, we reject all pixels with more than six counts and
  also containing more signal than contained in the surrounding 8 pixels summed.
  Using the aspect solution file (*sat*.fits), we determine the satellite pointing for
  each detection frame.  We then map the bad pixels in raw detector coordinates determined by {\tt xrtpipeline} and by
  our algorithm onto the sky on a frame-by-frame basis.  This is used to generate exposure maps for the
  full observation and as a function of time.

  Using the full exposure map, we determine the afterglow position centroid \citep[see,][]{butler07b} to fix the
  source extraction region.  We consider a 16 pixel radius source extraction region,
  surrounding by an annular background extraction region of outer radius 64 pixels.
  Running wavdetect \citep[see,][]{butler07b}, we then determine the positions of field sources
  in the image.  We mask out the regions corresponding to the field sources from the source
  and background extraction regions.  Also, using the Point Spread Function (PSF) model ({\tt swxpsf20010101v003.fits})
  at 1.3 keV, we determine the level of residual field source contamination in the
  source extraction region (typically negligible) for later subtraction.
  
  Initially ignoring pileup, we extract the source and background counts for each good time
  interval of data acquisition.  The fraction of lost signal and the scale factor relating the
  background in the source and background extraction regions is determined for each extraction
  using the time-dependent exposure map.  assuming these exposure corrections for the entirety of each time
  interval, we subdivide the counts in each interval so that a fixed signal-to-noise of 3 is
  achieved.

  In order to check and to account for pileup, we perform a coarse Bayesian blocking \citep{scargle98},
  with a strong prior weight against adding a new segment ($e^{-50}$).  Using the maximum observed count rate 
  in each segment thus determined, we find the minimum aperture necessary to reduce the source signal
  to levels where pileup is negligible.  The coarse blocking results in a small number of regions (typically 2--3)
  of differing inner extraction radius for an afterglow.  We assume pileup is important for count rates $>0.5$ cps 
  \citep[see also,][]{nousek06}.
  The light curves are verified to transition smoothly across regions of different inner extraction region radius.
  Using the time intervals and pileup corrected apertures thus determined, we rebin the data to 
  a signal-to-noise of 3 and recalculate the exposure correction for each time interval.  The
  final time regions and exposure corrections define our temporal extractions regions for the
  extraction of light curves in different energy bands and for extraction of spectra below.

\subsection{Windowed Timing (WT) Mode Light Curves}

  Our reduction of the WT mode data closely parallels our PC mode reduction, except that it is more natural
  to extract the WT mode data in raw detector coordinates than in sky 
  coordinates as done above for the PC mode data.  This is due to the readout mode; detector
  pixels are summed in RAW-Y and the resulting data are in column (RAW-X) format.

  Summing the data from each WT mode followup, we reject any RAW-X columns containing a $>10\sigma$ count
  rate relative to the background, after first ignoring pixels in the 16 pixel source extraction region.  We also reject
  any RAW-X columns containing 100 times more signal than the highest neighboring column (or $>100$ if the 
  neighbors contain no signal).
  Using the sky image determined from the PC mode data and the satellite aspect, we project the background
  onto the RAW-X axis and form a background mask for the 64 pixel outer radius and 16 pixel inner radius
  extraction region.  We do not allow masking of the pixels within the central 16 pixel source regions.
  If the source is bright ($>10^3$ cps), we recenter the source and background apertures.  Small aspect shifts $\sim$1
  pixel are not uncommon between the PC and WT mode data and must be accounted for.

  We determine the exposure corrections as for the PC mode data, but also adjusting the PSF model for the WT mode summing of RAW-Y pixels.
  We note that our exposure corrections account for source signal contained in the background region.
  We determine a pileup correction as above, but with a limiting source count rate of 150 cps \citep[see also,][]{nousek06}. 

\subsection{PC and WT Mode Spectra}

Spectral response files are generated using the {\tt xrtmkarf} task
for each time interval of interest.  Our invocation of the task ignores the exposure
maps calculated above, determining the energy dependence of of the source extraction assuming
only the inner and outer source and background regions.  We then adjust the normalization of the resulting Ancillary Response File
(ARF) to account for the actual loss in flux (0.5-8.0 keV) on a pixel by pixel basis using the divided time
intervals and associated exposure corrections determined above.
The spectra are fit in ISIS\footnote{http://space.mit.edu/CXC/ISIS/}.
For each spectral bin, we require a S/N of 3.5. 
We define S/N as the background-subtracted number of
counts divided by the square root of the sum of the signal counts and the
 variance in the background.   As done in Paper I, we restrict our attention to time-resolved spectra containing 500 or more counts
 or to spectra formed by grouping two or more of the 500 counts spectra.  

We fit the PC and WT mode data over the 0.3-10.0 keV range, also accounting for the systematic calibration uncertainties 
$\sim 3$\%\footnote{http://swift.gsfc.nasa.gov/docs/heasarc/caldb/swift/docs/xrt/spie05\_romano.pdf}.
In WT mode, we allow the detector gain to vary by $\pm 80$ eV\footnote{http://swift.gsfc.nasa.gov/docs/heasarc/caldb/swift/docs/xrt/xrt\_bias.pdf}.

\subsection{BAT Light Curves and Spectra}

We establish the energy scale and mask weighting for the BAT data by running the {\tt
bateconvert} and {\tt batmaskwtevt} tasks.  
The mask-weighting removes flux from background sources.  Spectra
and light curves are extracted with the {\tt batbinevt} task, and response
matrices are produced by running {\tt batdrmgen}.  We apply the systematic
error corrections to the low-energy BAT spectral data as suggested by the BAT
Digest website\footnote{http://swift.gsfc.nasa.gov/docs/swift/analysis/bat\_digest.html}, and fit the data using 
ISIS.  The spectral normalizations are corrected for satellite slews using the {\tt batupdatephakw} task.  For GRB~060124 below, BAT spectral fits are performed on the mask tagged light curve data in four channels, assuming the on-axis
response and also accounting for the systematic error.

\section{The Joint BAT$+$XRT Spectra of Three Events}
\label{sec:two}

There are two bright events in the XRT sample which
overlap in time entirely with what would commonly be thought of as
the prompt phase of GRB emission.
The observations by the XRT were made possible by a bright precursor
just minutes 
prior to each GRB observed in the BAT, on which the BAT triggered.
We therefore have both BAT and XRT data for each event, GRB~060124 and
GRB~061121.  We also discuss the bright event GRB~060614, which has excellent
XRT coverage due to an early, rapid spacecraft slew.

\begin{figure}[H]
\centerline{\includegraphics[width=3.5in]{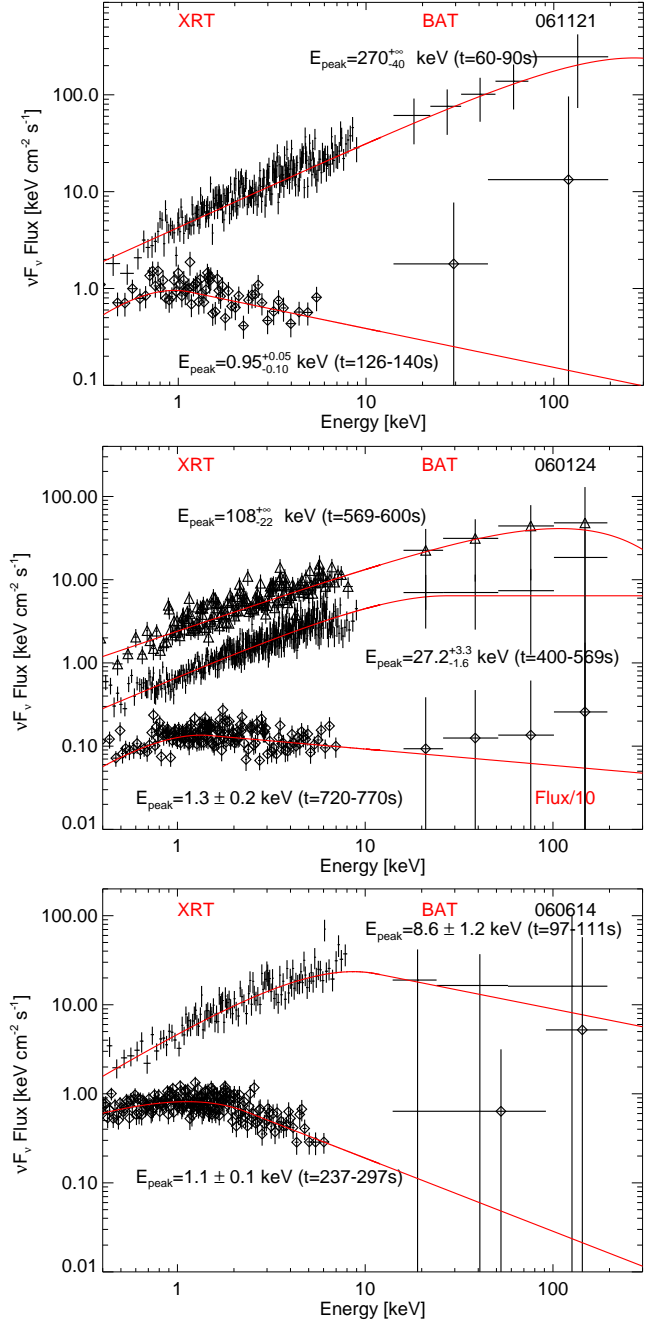}}
\caption{\small
Selected $\nu F_{\nu}$ spectra from GRBs 061121, 060124, and 060614, demonstrating the
Band model fits to a time varying spectral curvature --- as seen in plots
with logarithmic axes --- and $E_{\rm peak}$ evolution.
The X-ray data are corrected for photoelectric absorption using the
best fit late-time values of $N_H$ in Figures \ref{fig:3main_spec_panel} (left).
The softest spectrum in the middle panel is divided by a factor ten for legibility.
The counts spectra are jointly fit by forward folding the Band model through
the instrument response matrices.  For the spectral fits (Table 1), the BAT data are not binned as shown here.}
\label{fig:nuFnu_plots}
\end{figure}

Figure \ref{fig:nuFnu_plots} displays spectral fits to a selected set of time-resolved intervals
in each events.  The best-fit model parameters are given in Table 1.  The time evolution
of these parameters are presented and discussed in detail in the next three subsections.

\subsection{GRB~060124}
\label{sec:060124}

Swift-BAT triggered and located the precursor to GRB~060124,
allowing the XRT to slew and begin simultaneous observations
106s later \citep{holland06}.  This event is also discussed in
\citet{romano06b}.  The 0.3-10.0 keV light curve is plotted in
Figure \ref{fig:060124_lc}.  There are two prominent peaks.  
As shown in the background (lighter two shades of gray), the time profile in 
the soft XRT channel (0.3--1.3 keV) is broader than that in the
hard channel (1.3--10.0 keV).  The BAT light curve shows even
narrower time structure and resolves the broad first XRT peak into
at least 3 sub-peaks.  The light curve after the flare ($t>10^4$s) and
extending to 22 days is 
well fit by a powerlaw $t^{-1.32\pm0.01}$ ($\chi^2/\nu=535.2/465$).

\begin{figure}[H]
\centerline{\includegraphics[width=6.0in]{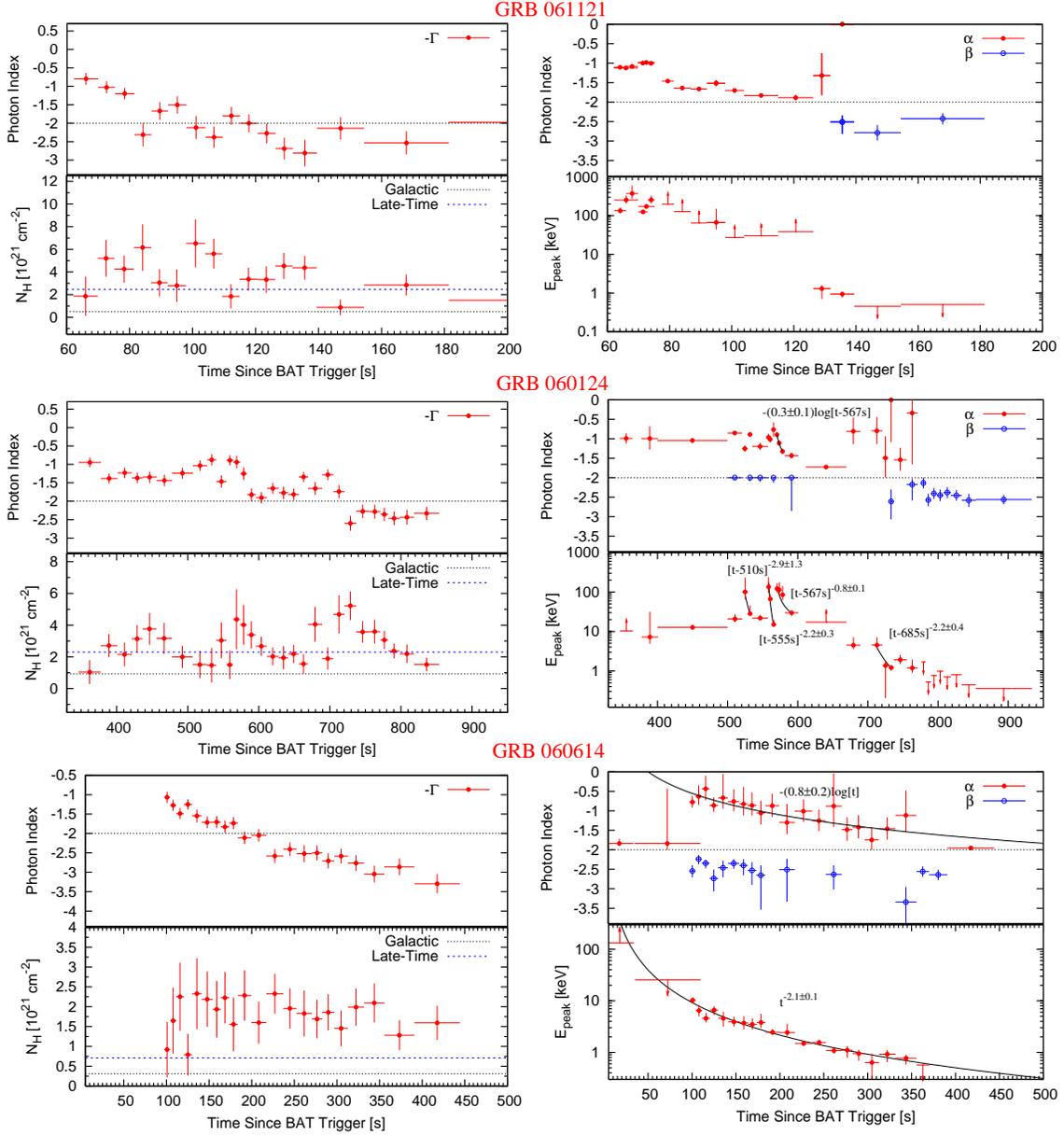}}
\caption{\small
Powerlaw (left panels) and Band model (right panels) fits to the GRBs 061121,
060124, and 060614.  Time-correlated $N_H$--$\Gamma$ variations in the left plots are
better modelled by spectral models with time-evolving $E_{\rm peak}$'s in the right plots.
The $N_H$ values peak when $E_{\rm peak}\approx 1$ keV.
The powerlaw fits are performed for only the X-ray data, whereas
the Band fits (actually nested powerlaw then exponential times powerlaw then Band fits,
as described in the text) apply to the X-ray and $\gamma$-ray data.  
Trends in the Band model parameters, when observed,
are fitted and presented in Table 3 and in the text.
These time variations are given relative to
the approximate pulse start times.
Galactic column densities are taken from \citet{dickey1990}  The Band fits use the
late time $N_H$ values plotted in the left panels, derived from X-ray fits at $t>10^4$s.}
\label{fig:3main_spec_panel}
\end{figure}

We group the XRT data into $\gtrsim 500$ counts spectra
and fit powerlaws (Figure \ref{fig:3main_spec_panel}, left).  Each fit is
statistically acceptable, with a reduced $\chi^2$ of order unity.  The
photon index $\Gamma$ is observed to decrease in time, although with
modulation in time that correlates with the X-ray flux and with $N_H$
(see explanation in Section \ref{sec:glob}).
At late times ($t>10^4$s), the $N_H$ values asymptote to the blue, dashed
curve ($N_H=2.3 \pm 0.2 \times 10^{21}$ cm$^{-2}$) plotted in the figure.

To study the time varying log-log curvature, we jointly fit the BAT and XRT data
using the \citet{band93} model.  Here, we choose extraction regions
which allow for a BAT signal-to-noise of 20 or higher.  We also
fix the column density $N_H$ to the late time value.  The model fit
is actually a progression of fits of nested models \citep[e.g.,][]{protassov02}, from the 
simplest powerlaw model to a powerlaw times exponential model, to the 
smoothly broken powerlaw
Band model.  Each more complex model has one additional degree of
freedom.  We accept or reject the more complex model at each stage by
requiring $\Delta \chi^2>2.706$ (i.e., 90\% confidence).  If the data
are acceptably fit by only the powerlaw model, we quote a limit on
$E_{\rm peak}$ using either the exponential times powerlaw model
(for $\Gamma<2$) or the constrained Band formalism 
\citep[][; for $\Gamma>2$]{sakamoto03}.  
In order that $E_{\rm peak}$ correspond to a peak in
the $\nu F_{\nu}$ spectrum, we require the low energy index
$\alpha>-2$ and the high energy index $\beta<-2$.
After finding that the fits were consistent with $\alpha<0$, as also
found for {\it BATSE}~GRBs \citep{preece00,kaneko06}, we included this 
as a constraint to derive the tightest error bounds on the other model parameters.  

As shown in Figure \ref{fig:3main_spec_panel} (right), the data are better fit 
($>90$\% confidence) with the Band model in most of the time regions.  
The peak energy rises and declines
with each of the four prominent light curve pulses.  For each pulse,
we present powerlaw fits to the $E_{\rm peak}$ declines.  The rises are
not well measured, as is also typically the case for {\it BATSE}~bursts
\citep[e.g.,][]{krl03}.
Prior to $t\lessim 800$s, the observed spectrum
corresponds mostly to the low energy portion of the Band
model spectrum, except episodically at the flare troughs, where
$E_{\rm peak}$ enters the X-ray band.  These times regions are also
those of highest $N_H$ in Figure \ref{fig:3main_spec_panel} (left).  The third pulse
decline exhibits a strong evolution in both $\alpha$ and $E_{\rm peak}$.
After $t\sim 800s$ the observed spectrum corresponds to the high
energy portion of the model spectrum, and $E_{\rm peak}$ has transited
the X-ray band.  Figure \ref{fig:nuFnu_plots} (middle) plots the $\nu F_{\nu}$ spectrum
at 3 time epochs.

Motivated by the watershed event GRB~060218 \citep[Paper I;][]{campana06a}, we also attempt to
fit the X-ray curvature using a powerlaw plus blackbody model.  The
fits to the X-ray data alone are provocative and show a smooth
temperature decline after each of the two major pulses.  However, the
fits are statistically unacceptable when we also attempt to account for
the BAT data.  This is also true for the GRB~061121 spectra discussed in
the next sub-section.  
This should be taken as a caveat also to the powerlaw plus blackbody fits
presented for the XRT data in Paper I, where the derived blackbody
temperature variation may imply instead to $E_{\rm peak}$ variations.
We note, however, that the X-ray spectra of the unusual GRB~060218 burst and afterglow are better
fit by a blackbody plus powerlaw than by a Band model (Paper I).
We do not consider the possibility of two powerlaws and
a blackbody for the bursts discussed here.  

\subsection{GRB~061121}
\label{sec:061121}

\begin{figure}[H]
\centerline{\includegraphics[height=3in]{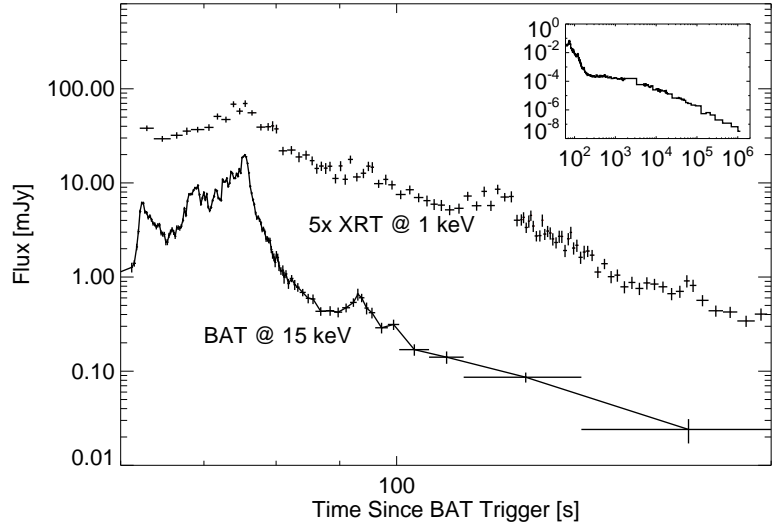}}
\caption{\small
The hard X-ray BAT and XRT light curves for GRB~061121.  The late-time
light curve (a plateau at $t\sim 200$s followed by a decline beginning
at $t\sim 3000$s) is plotted in the sub-panel.  The XRT light curve
has been multiplied by 5 to bring it above the BAT light curve.}
\label{fig:061121_lc}
\end{figure}

\swift-BAT triggered on and began observing the precursor to GRB~061121
55s prior to the XRT slew toward and onset of the main GRB event
\citep{page06}.  The \swift~team designated this event a
``Burst of Interest'' \citep{gehrels06} due to the rare simultaneous
detections in the BAT and XRT bands and at longer wavelengths.
As shown in Figure \ref{fig:061121_lc}, the $\gamma$-ray and X-ray
light curves show multiple peaks, with most of the prominent time structure
apparent in only the $\gamma$-ray band.

In Figures \ref{fig:3main_spec_panel} (top), we show the results of powerlaw
and Band model fits to the 061121 data.  The data are not of
as high signal-to-noise in the X-ray band as the 060124 data, however,
many of the same trends are apparent.  There is a hard to soft evolution
apparent in the powerlaw photon index and a correlation between the
photon index and $N_H$.  The Band model photon index
goes from the low energy side to the high energy once $E_{\rm peak}$
has crossed the X-ray band.  $E_{\rm peak}$ also appears to rise and
fall with flaring prior to 80s.  The $\nu F_{\nu}$ spectrum
is plotted at two epoch in Figure \ref{fig:nuFnu_plots} (top).
For the Band fits, we use the late
time ($t>10^4$s) $N_H=2.5 \pm 0.3 \times 10^{21}$ cm$^{-2}$.

{\it XMM} data for this event beginning after $t\approx 6$hrs show
consistent powerlaw fits with our late-time fits \citep{butleretal07}.
In particular,
$N_H = 1.71^{+0.03}_{-0.02} \times 10^{21}$ cm$^{-2}$, consistent with
our late-time $N_H$ at the $2\sigma$ level and well below the mean early-time 
value.  {\it XMM}~and XRT data generally agree well with respect to the
late-time $N_H$ determinations \citep[e.g.,][]{moretti06}.

\subsection{GRB~060614}
\label{sec:060614}

\begin{figure}[H]
\centerline{\includegraphics[height=3in]{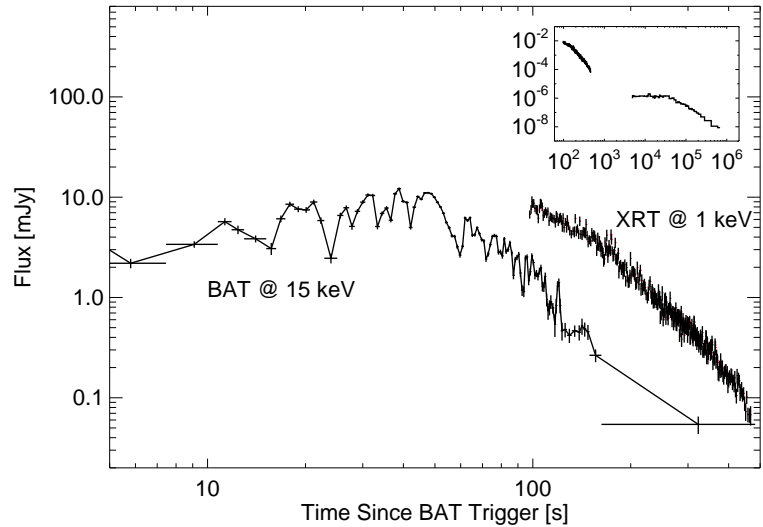}}
\caption{\small
The hard X-ray BAT and XRT light curves for GRB~060614.  The late-time
light curve (a plateau at $t\sim 200$s followed by a decline beginning
at $t\sim 4000$s) is plotted in the sub-panel.}
\label{fig:060614_lc}
\end{figure}

The GRB~060614 \citep{parsons06} afterglow
fades rapidly as a powerlaw from the prompt emission, with no flaring (Figure \ref{fig:060614_lc}).
There is excellent
BAT$+$XRT coverage during the prompt tail emission lasting to $t\sim 150$s.  
We observe weak $N_H$--$\Gamma$ correlated modulations during the rapidly fading tail,
which would imply an $N_H$ that decreases in time, reaching the value marked
by the dashed line in Figure \ref{fig:3main_spec_panel} (left) by $t=10^4$s.

However, the Band model fits show an $E_{\rm peak}$ which passes through
the X-ray band without requiring a varying $N_H$.  Extrapolating backward through the prompt emission, the best
fit decay also fits two $E_{\rm peak}$ limits derived for the BAT only prompt emission.
Figure \ref{fig:nuFnu_plots} (bottom) plots the $\nu F_{\nu}$ spectrum at two epochs.
Expressed in terms of the flux $F_{\rm XRT}$ as measured by the XRT rate, $E_{\rm peak} \propto F_{\rm XRT}^{-0.72\pm0.03}$.

The low-energy photon index $\alpha$ also appears to evolve in time after the main GRB emission.

\subsection{Hardness Plots for GRBs 061121, 060124 and 060614}
\label{sec:hardness_plots}

It will be useful below to see how the spectral evolution in the early 
X-ray light curves of GRBs 061121, 060124 and 060614 impacts the X-ray
hardness ratio.  We define this is the ratio of counts in the
1.3--10.0 keV band to the counts in the 0.1--1.3 keV band.   The average
hardness ratio ($HR$) for most afterglows is 1.

\begin{figure}[H]
\centerline{\includegraphics[width=7.0in]{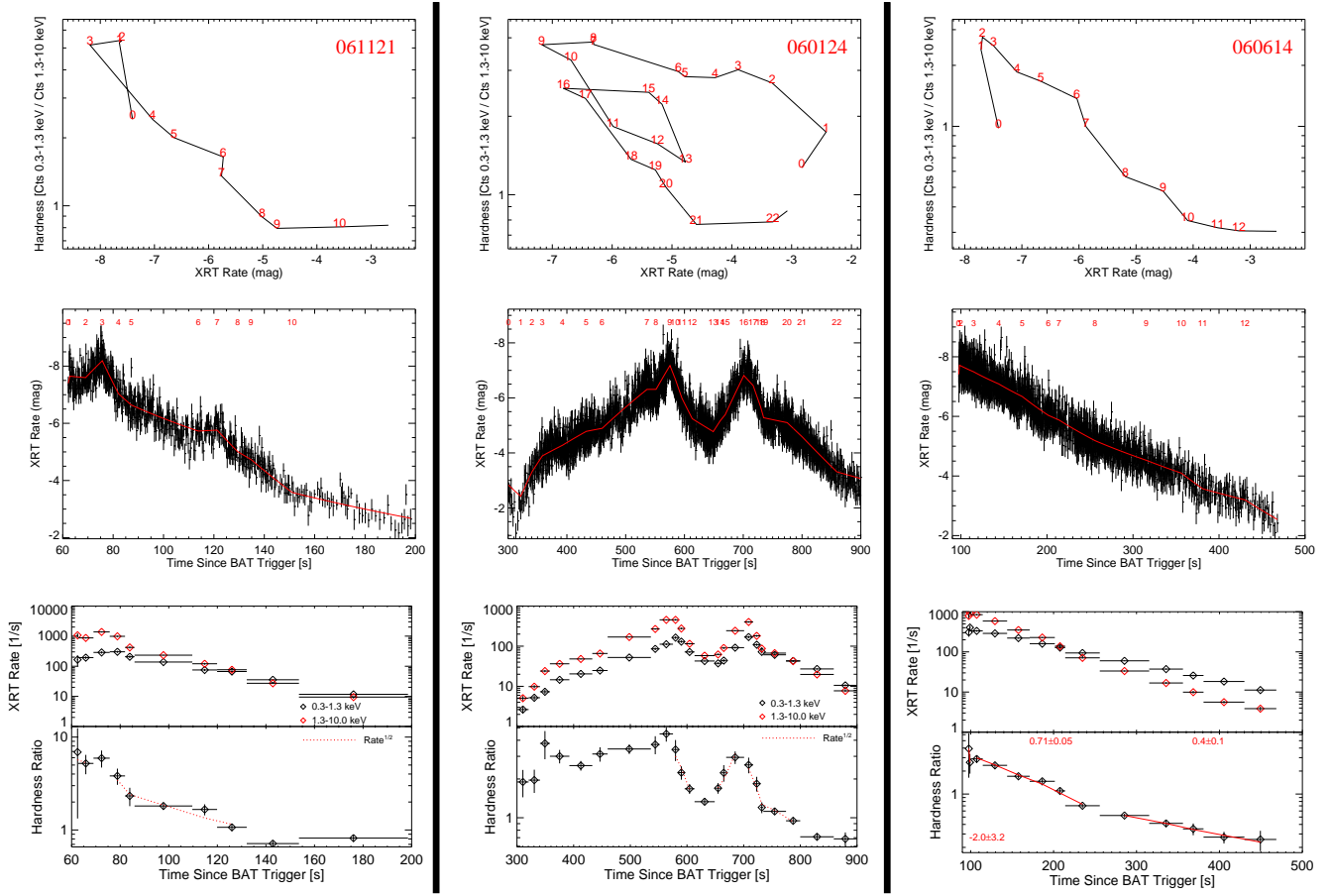}}
\caption{\small
The hardness evolution in GRBs 061121, 060124, and 060614.  (top plots) Hardness versus
rate fit, indexed as a function of time, showing evolution along roughly 
parallel tracks.  (middle plots) The X-ray light curve and
fit (red curve) as source of the time indexing.  (bottom plots) The X-ray light 
curve in each band (hard is red, soft is black)
for each time segment and the hardness during each time segment.  This
is well fit during the declines by the square-root of the rate (dotted
red line) in GRBs 061121 and 060124 and by a power close to the square root
of the rate for GRB~060614.}
\label{fig:main_hard_panel}
\end{figure}

Figure \ref{fig:main_hard_panel} in 9 panels shows the hardness and rate
time profiles for GRBs 061121, 060124, 060614.  The middle panels (looking top to bottom) show the X-ray
light curve fit using an extension of the Bayesian blocks algorithm
\citep{scargle98} to piecewise logarithmic data.  The rate and hardness data are
fit jointly, allowing the minimum number of powerlaw segments such
that $\chi^2/\nu \sim 1$.  The fits to the rate and hardness are plotted
in the top and middle panels, indexed according to time.
The hardness tracks the flux 
and moves along roughly parallel tracks.  In the bottom panels,
the flux in both XRT bands (top panel) and the hardness (bottom panel)
are plotted for each powerlaw segment.  During the decline phase of
each pulse, the hardness scales as the square-root of the rate for GRBs 061121 and 060124.
For the GRB~060614, the hardness and flux track as found above for $E_{\rm peak}$ and flux.

Each pulse in GRB~060124 peaks at roughly the same time, independent of energy band.
There is, however, a hardness rise during the flux rise because the hard
band increases more rapidly.  There is also a modest overall hard to soft trend
throughout the light curve.

The hardness plot does not capture the strong spectral variations 
between 500 and 600s in GRB~060124, which are apparent from the broad band fits
(Figure \ref{fig:3main_spec_panel} middle) and occur mostly for $E_{\rm peak}$
above the XRT bandpass.  The time dependences of $E_{\rm peak}$ during
this region and later are given in the figure.  The $E_{\rm peak}$
dependence can also be given in terms of the flux $F$, in order to sidestep
the problem of unknown start time.  
For all but the last flare, where we use the XRT
count rate, we use the BAT 15--350 keV count rate for the flux.
For pulses 1--4, we find
$E_{\rm peak} \propto F_{\rm BAT}^{-3.6\pm1.7}$, $F_{\rm BAT}^{-1.8\pm0.5}$, $F_{\rm BAT}^{-0.3\pm0.1}$,
$F_{\rm XRT}^{-1.2\pm0.2}$.  
In the bottom right left panels of Figure \ref{fig:main_hard_panel}, we
show that the hardness can be described by the square root of the
observed flux, as is common for GRBs at higher energies observed with {\it BATSE}~
\citep[e.g.,][Section \ref{sec:discussion}]{bnr01,rnp02,krl03,ryde05}.

For GRB~061121, the hardness plots show
an initial hardening followed by a decrease in the hardness which scales
well with the square root of the X-ray rate.  There may be broad pulses
on top of the decline, although these have only
a minor impact on the hardness.  GRB~060614 appears to mostly to exhibit a secular decline
in both flux and hardness, corresponding to the fading tail of the prompt emission.

For each GRB, the hardness plot capture the $E_{\rm peak}$ evolution
in general terms.  Both ($HR$ and $E_{\rm peak}$)
decrease during rate declines at a similar power of the rate.  It is 
apparently not possible to cleanly if at all separate evolution
of $\alpha$ from evolution of $E_{\rm peak}$, given the hardness alone.
From Figures \ref{fig:3main_spec_panel} (top right) and \ref{fig:3main_spec_panel} (middle right) and also from time-resolved spectral
studies of many GRBs (Section \ref{sec:discussion}), these parameters tend to evolve simultaneously.

\section{Example Spectra for 4 Other Early Afterglows}
\label{sec:others}

\begin{figure}[H]
\centerline{\rotatebox{270}{\includegraphics[width=2.3in]{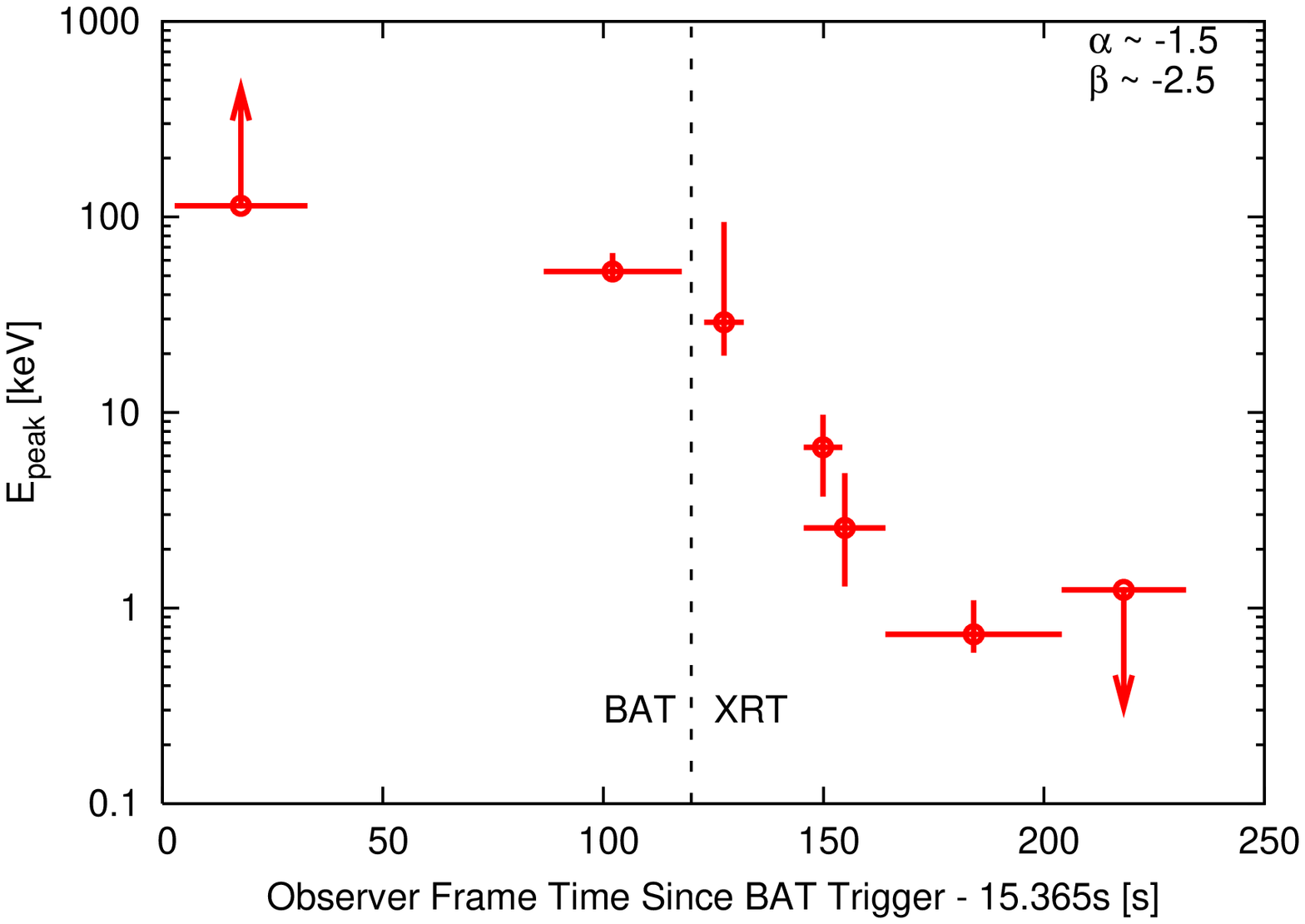}}}
\centerline{\includegraphics[width=3.6in]{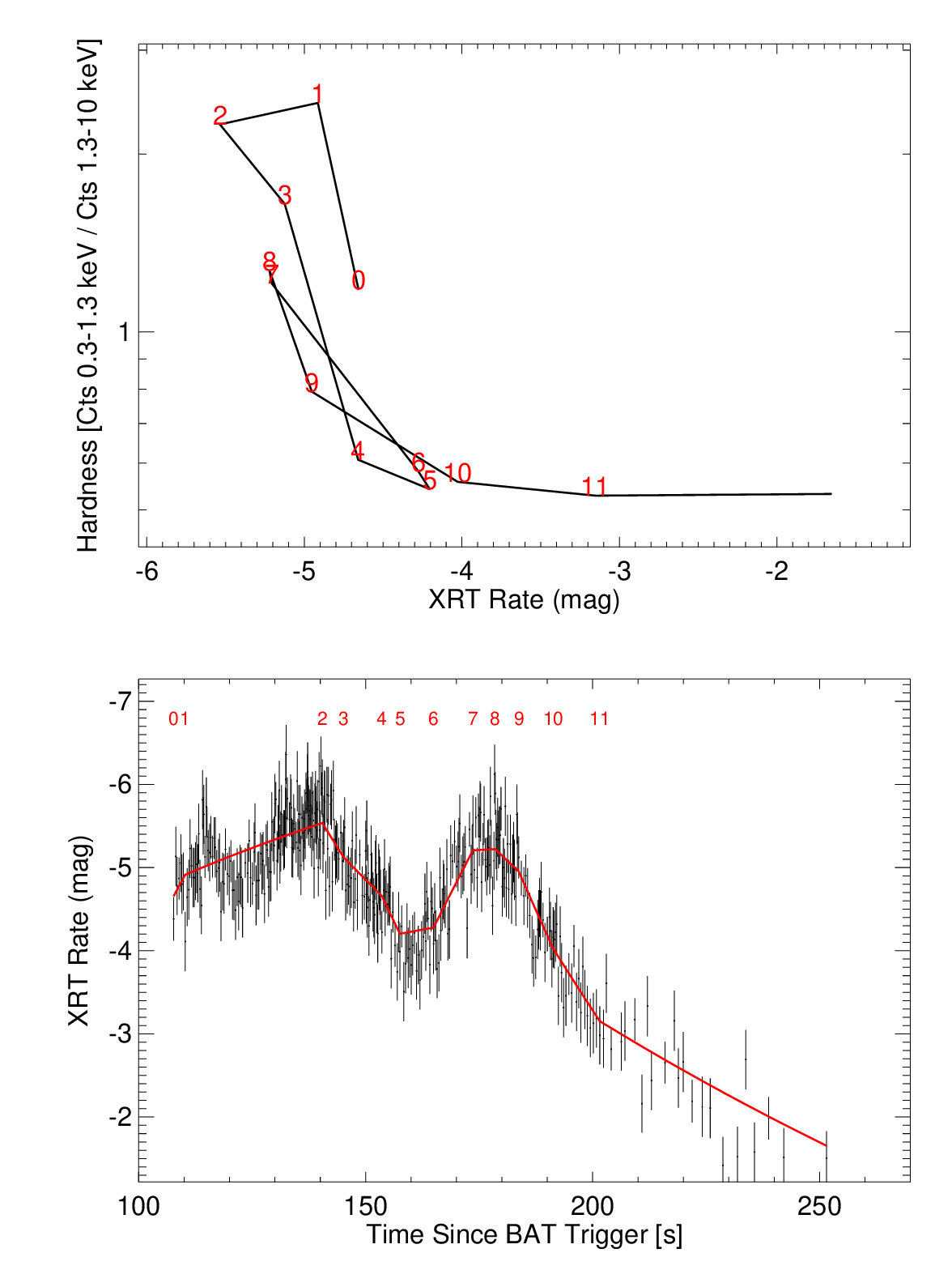}}
\caption{\small
Plots of the hardness and $E_{\rm peak}$ evolution for flares after GRB~060714.  The
Band fits allow only a coarse time resolution, whereas the hardness study demonstrates
a fine timescale changes in the spectrum which track the flux across flares.
$E_{\rm peak}$ evolution
from Band model fits to the BAT and XRT data (top plot).
Typical values
for the photon indices ($\alpha$,$\beta$) are given.
Hardness versus
rate fit (middle and bottom plots), indexed as a function of time, showing parallel evolution
tracks.}
\label{fig:060714_both}
\end{figure}

Most early X-ray afterglows have a low signal-to-noise or no
coincident detection by the BAT.  It is possible to derive
$E_{\rm peak}$ values or limits for these early on, given the BAT
data.  Late time $E_{\rm peak}$ from the X-ray data typically show
values in or passing through the XRT band after one to several minutes.
The spectral evolution from one such event, GRB~060714
\citep{krimm06}, is shown in 
Figure \ref{fig:060714_both}.

The hardness plot (Figure \ref{fig:060714_both} middle) allow for a finer
time sampling of the spectra evolution.  The hardness (likely also 
$E_{\rm peak}$) rises and declines with the flux along the same track
in the hardness--rate plane as two flares take place.  The column
density (not plotted) is a factor two larger in the time interval
140--170s than outside that interval, indicating an $E_{\rm peak}$ passage.
There are a handful of examples with higher signal-to-noise XRT observations.

\begin{figure}[H]
\centerline{\includegraphics[width=5in]{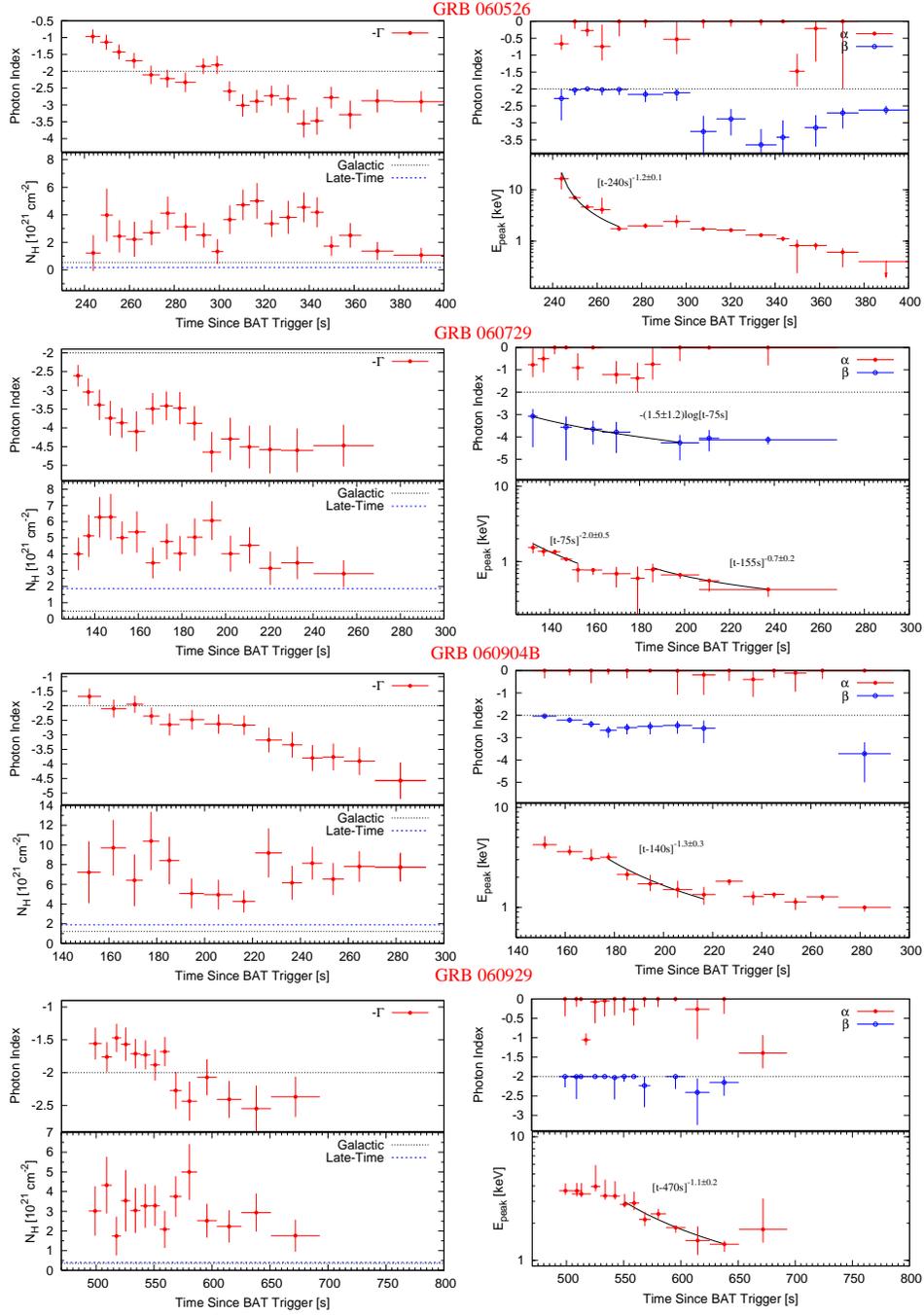}}
\caption{\small
Powerlaw (left panels) and Band model (right panels) fits to the GRBs 060526,
060729, 060904B, and 060929.
Time-correlated $N_H$--$\Gamma$ variations in the left plots are
fit by spectral models with time-evolving $E_{\rm peak}$'s in the right plots.
Trends in the Band model parameters, when observed,
are fitted and presented in Table 3 and in the text.
See also Figure \ref{fig:3main_spec_panel}.
In the Band model plots, $\alpha$ values which appear to be above and outside of
the plotted range are those which reach and remain at the paramater bound $\alpha=0$
 (see Section \ref{sec:060124}).}
\label{fig:4back_spec_panel}
\end{figure}

\begin{figure}[H]
\centerline{\includegraphics[width=7in]{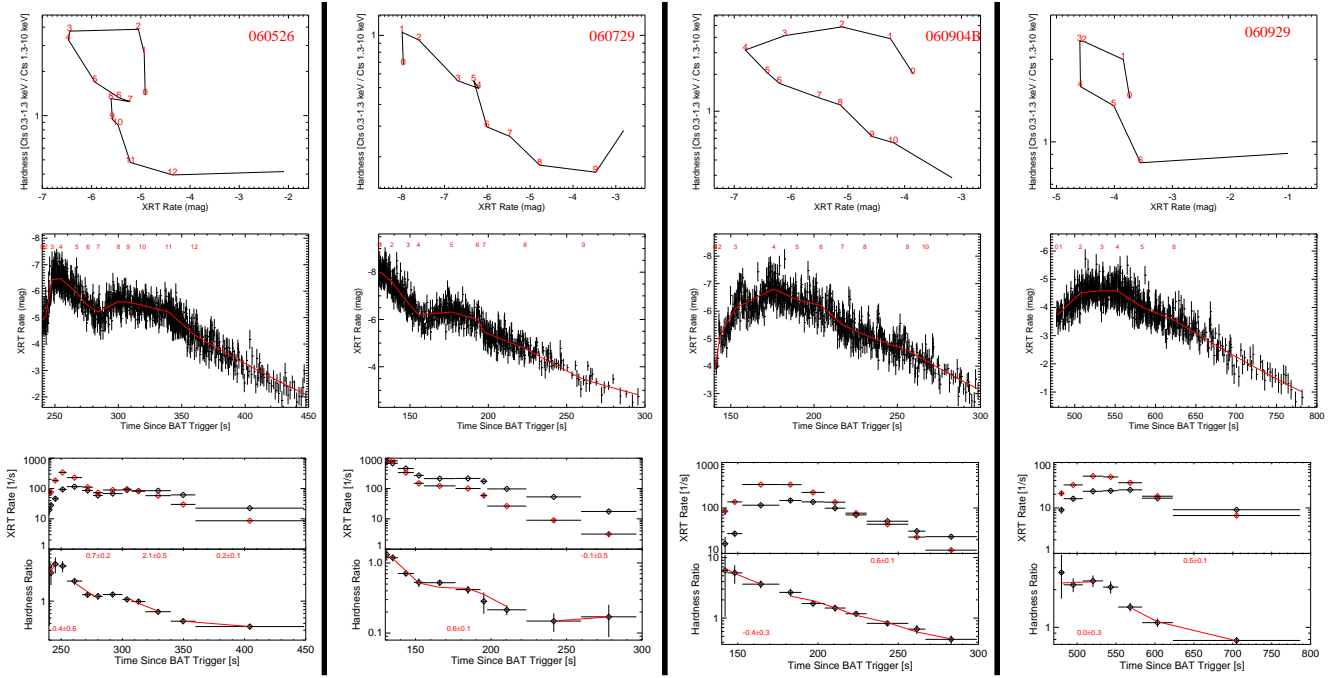}}
\caption{\small
Hardness plots for GRBs 060526, 060729, 060904B, and 060929.  
(top plots) Hardness versus
 rate fits, indexed as a function of time, showing evolution along roughly 
 parallel tracks.  (middle plots) The X-ray light curve and
 fit (red curve) as source of the time indexing.  (bottom plots) The X-ray light 
 curve in each band (hard is red, soft is black)
 for each time segment and the hardness during each time segment.  
   This hardness is well fit during the declines 
   by the rate to a power close to 0.5 (dotted red curves).
 See also Figure \ref{fig:main_hard_panel}.}
\label{fig:back_hard_panel}
\end{figure}

The GRB~060526 \citep{campana06b} afterglow
exhibits time-correlated $\Gamma$--$N_H$ variations and a corresponding
rapid
then smooth decline of $E_{\rm peak}$ through the XRT band
(Figure \ref{fig:4back_spec_panel} left).
The initial GRB pulse ($t<9.4$s) is well fit by a simple
powerlaw ($\alpha=-1.6\pm 0.2$, $\chi^2/\nu=16.83/16$), and
we derive $E_{\rm peak}>80$ keV (90\% confidence).
The flare at $t\sim 250$s is detected by the BAT as well, and we use
the BAT data to obtain the best Band model fits.
The Band model photon indices are poorly measured.
The composite flare and decline is shown in Figure \ref{fig:back_hard_panel}.
The hardness evolves similarly to the best-fit $E_{\rm peak}$ values.

The very bright afterglow to GRB~060729 \citep{grupe06a} continues
to be detected 4.5 months after the GRB.
The GRB is over and done with by $t\sim 130$s in the BAT. 
We find $E_{\rm peak}>50$ keV (90\% confidence).
After $t>100$s in the XRT, there is a rapid decline, interrupted by
a flare or rise at 160s (Figure \ref{fig:back_hard_panel}).  Time-correlated
$N_H$--$\Gamma$ variations and an $E_{\rm peak}$ passage through the
X-ray band are similar to those discussed above (Figure \ref{fig:4back_spec_panel}).  We observe that
$E_{\rm peak}$ declines with the X-ray rate as $F_{\rm XRT}^{-0.4\pm0.1}$
both before and after the mild flare at $t\sim 180$s.  There is also
a possibly significant decline in $\beta$ with time.

The hardness declines
by an order of magnitude, reaching a minimum at $t\sim 250$s,
and then increases to the late time ($t>10^3$s) value.  
Note that no clear coincident change is present in the rate plot.
The hardness plot demonstrates that the late time emission is spectrally
different from the early emission and that its onset occurs at $t\sim 250$s.

Modest but clear $N_H$--$\Gamma$ variations are seen for
GRB~060904B \citep{grupe06b}.  
The prompt emission ($t<8.3$s) has
$E_{\rm peak} = 125^{+135}_{-30}$ keV.  $E_{\rm peak}$ transits the X-ray
band nicely (Figure \ref{fig:4back_spec_panel}).  The hardness evolution 
shows the usual time dependence
in the declining tail of the flare (Figure \ref{fig:back_hard_panel}).
$E_{\rm peak}$ decays versus the rate as $F_{\rm XRT}^{-0.7\pm0.2}$.

The emission for GRB~060929 \citep{markwardt06} at $t<13$s
exhibited $E_{\rm peak}>75$ keV.  The X-ray flare peaking at $t\sim 550$s
is weakly detected by the BAT.  In the XRT, there is a clear softening
trend (Figure \ref{fig:back_hard_panel}), likely $N_H$--$\Gamma$ variations,
and an $E_{\rm peak}$ declining through the X-ray range 
(Figure \ref{fig:4back_spec_panel}).  $E_{\rm peak}$ drops with the
X-ray rate as $F_{\rm XRT}^{-0.6\pm0.1}$.  The hardness reaches a minimum
at $t=630\pm10$s.

\section{Discussion}
\label{sec:discussion}

\subsection{Global Sample Properties}
\label{sec:glob}

\begin{figure}[H]
\centerline{\includegraphics[width=4.0in]{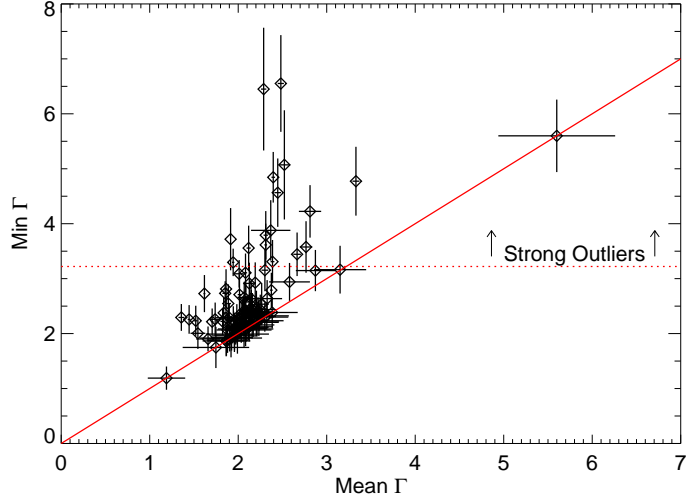}}
\caption{\small
As also discussed in Paper I, there is an outlier population of very soft
\swift~XRT afterglow time regions with respect to the majority population 
clustering near photon index $\Gamma \sim 2$.}
\label{fig:gammas}
\end{figure}

In terms of the spectral evolution properties, we see no apparent difference
between the fading tales of flare-like X-ray emission and the rapid X-ray
declines often observed to trail flaring in the BAT
\citep[e.g.,][]{tag05,bart05,cus06,vaughan06}.  Indeed, based solely
on timing properties, many of the rapid declines also appear to
have superimposed flaring (e.g., 060729, Figure \ref{fig:back_hard_panel};
061121, Figure \ref{fig:main_hard_panel}).  
The rapid declines are thought to be the fading tail of the prompt
emission \citep{pain06,yama06,lnb06,zhang06}, and the X-ray flares are thought
to be due to later central engine activity \citep{zhang06,ioka05,fan05}.
We observe a clear distinctions between the spectra measured before the
light curve plateau and after the start of the plateau; only the late
spectra exhibit a tight clustering
with $\Gamma\approx 2$ (Figure \ref{fig:gammas}; Paper I; \citet{butNkoc07}).

\begin{figure}[H]
\centerline{\rotatebox{270}{\includegraphics[width=3.0in]{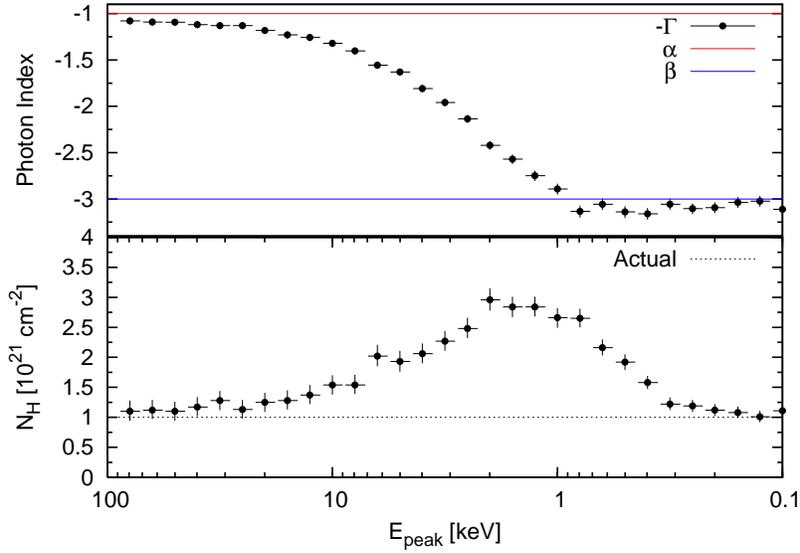}}}
\caption{\small
Powerlaw fits to high signal-to-noise data ($10^4$ counts, 0.3-10.0 keV)
simulated from a Band spectral model with $\alpha=-1$ and $\beta=-3$.
Each fit is statistically acceptable ($\chi^2/\nu\sim 1$).  With
the passage of the $\nu F_{\nu}$ peak energy $E_{\rm peak}$, the best-fit
photon index $\Gamma$ steepens smoothly.  An artificial increase
in the inferred X-ray column density $N_H$ linearly proportional to $\Gamma$
is observed for peak energies $E_{\rm peak}$ in the XRT bandpass. The
effect is present, with larger $N_H$ error bars, for spectra with few
counts.}
\label{fig:sim_spec}
\end{figure}

\begin{figure}[H]
\centerline{\includegraphics[width=4.0in]{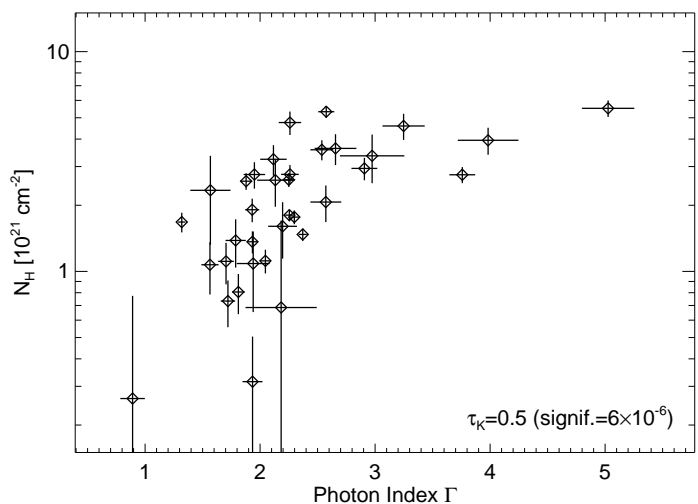}}
\caption{\small
Time integrated spectral fits to the flares in Table 2, also shown in Figure 
\ref{fig:hard_flares}, demonstrate a significant positive correlation 
between the column density parameter $N_H$ (observed minus Galactic)
and the photon index $\Gamma$.
Although these quantities are correlated for a given spectrum, we do
not expect a correlation at different times for the same event (see below) 
or at any time for separate events as found here.  This is evidence tying
the X-ray flares to an excess spectral curvature at X-ray wavelengths.}
\label{fig:nh_vs_gamma}
\end{figure}

Figure \ref{fig:sim_spec} shows what we expect to measure from powerlaw fits
to a time-evolving Band model spectrum.
As $E_{\rm peak}$ enters the X-ray band, the spectral curvature as would be seen
on a plot with logarithmic axes
increases and the inferred X-ray column density increases linearly with an increasing
inferred photon index $\Gamma$.  This occurs despite the fact that only
$E_{\rm peak}$ changes in the simulation.  
Figure \ref{fig:nh_vs_gamma} suggests that the effect is common in the XRT data
(Section \ref{sec:nh}).

Figure \ref{fig:hard_flares} (left) shows that the
flares (Table 2) and rapid X-ray declines exhibit
significant hardness--intensity and hardness--fluence correlations which match closely the 
correlations observed for GRBs (Section \ref{sec:xprompt} below).

\begin{figure}[H]
\centerline{\includegraphics[width=6in]{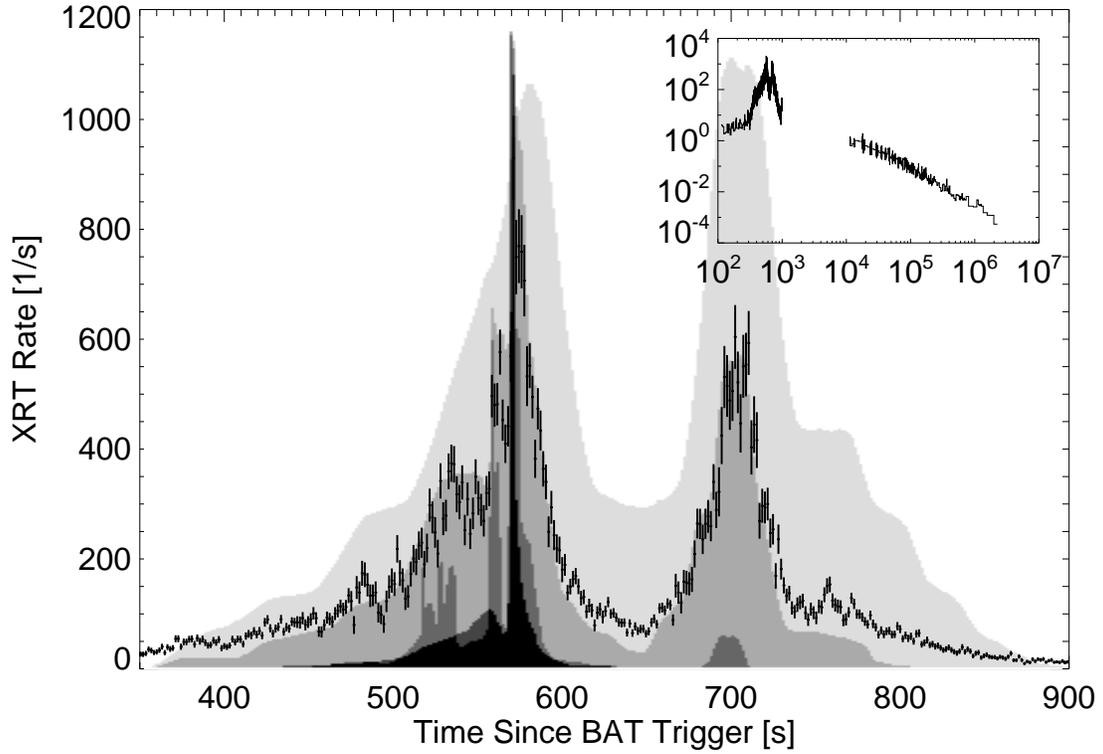}}
\caption{\small
The light curve for GRB~060124.  The X-ray data (0.3-10.0 keV) are plotted
in black.  The shaded regions in the background depict the X-ray
light curves in two energy bands (0.3--1.3 keV and 1.3--10.0 keV) and
in the hard X-ray/$\gamma$-ray bands of BAT (15--100 keV and 100--350 keV).
The background light curves are each denoised and normalized to their peak 
intensity.  The harder regions are darker.  The sub-panel shows the early
and late XRT light curve.}
\label{fig:060124_lc}
\end{figure}

\begin{figure}[H]
\centerline{\includegraphics[width=7.0in]{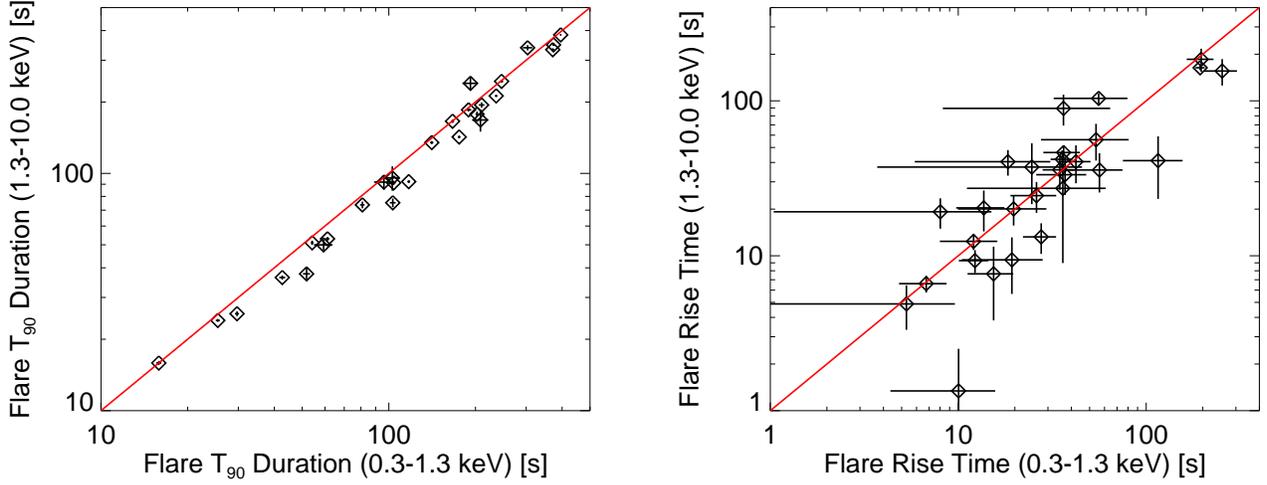}}
\caption{\small
Timing statistics for the bright flares in (Table 2).
The flare $T_{90}$ durations (left plot) and rise times (right plot)
are systematically longer in the soft X-ray
channel (left plot), by $8\pm1$\% and $25\pm5$\%, respectively.  \citet{norris96,fenimore95,fenimore96} discuss
similar properties of GRB pulses.}
\label{fig:durations}
\end{figure}

For GRBs it is common to observe finer time structure at higher energies
as compared to low energies \citep{norris96,fenimore95,fenimore96}.  Pulses tend to be narrower, fade
more rapidly, and evolve stronger spectrally at high energies.  Consistent with this, the X-ray flares (and
also the rapid declines) appear longer ($8\pm 1$\% on average, Figure \ref{fig:durations} left)
and with smoother time structure (e.g., Figure \ref{fig:060124_lc}) at softer energies.  
This can be understood as the effect of $E_{\rm peak}$ evolving into the X-ray band,
which allows the X-ray emission to be observed for longer \citep[e.g.,][and Section \ref{sec:interp}]{krl03}.
Although it is difficult to see by a eye, there is also evidence for a $25\pm 5$\% increase in the flare rise
time with decreasing X-ray energy band (Figure \ref{fig:durations} right).
This is close to the expected pulse broadening fraction from an extrapolation of the GRB behaviour,
$1-(1.3/0.5)^{-0.4}\approx 30$\%, where 0.5 and 1.3 keV are used as approximate lower bandpass energies.
Given the possibility that resolved $\gamma$-ray flares are blurred together in the X-ray band
(e.g. Figure \ref{fig:060124_lc}), however, it is not clear how meaningful this apparent consistency is.

\subsection{The Physical X-ray Column Density Does Not Vary}
\label{sec:nh}

The time-resolved XRT afterglows are well fit by absorbed powerlaws at all epochs
(see also Paper I).   Prior to a characteristic hardness variation turn-off time $T_H\approx 10^2-10^4$s,
which we discuss for a large sample of bursts in \citep{butNkoc07}, there is strong evolution in both
the best-fit photon indices and the best fit column densities $N_H$.  After this time,
the quantities typically do not vary.  To fit more complicated models to the early
time afterglows, we have found it necessary to jointly fit the BAT and XRT data
(when possible) and to tie the column density to the value measured at late time.
The late time value is typically not the Galactic value.

Band model fits are able to account for both the BAT and XRT emission without a time
variable column density \citep[see, also,][]{falcone06}.  The ubiquitous hardness evolution appears to be best understood
in terms of an evolving $E_{\rm peak}$, as we discuss in detail below.

\begin{figure}[H]
\includegraphics[width=6.5in]{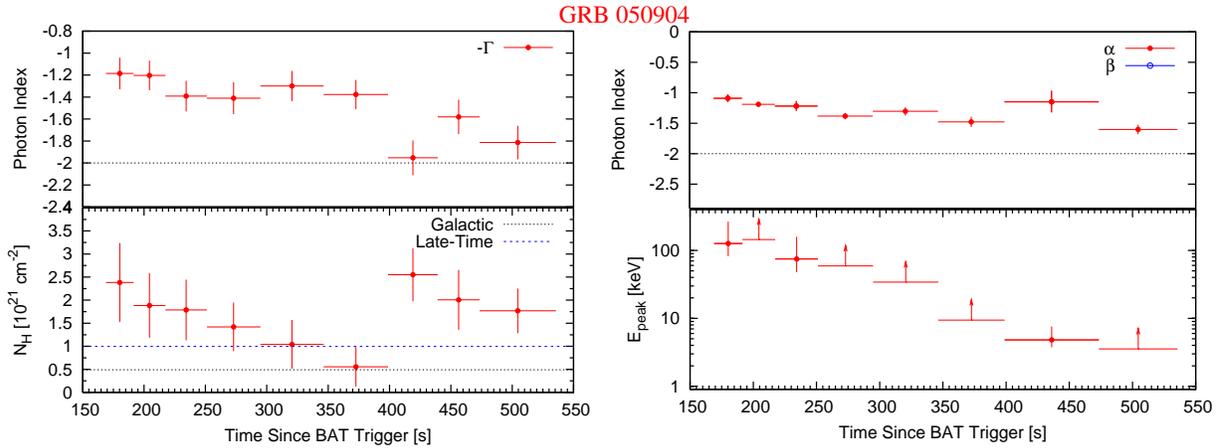}
\caption{\small
We believe $N_H$ variations are an incorrect explanation for the spectral evolution in the flaring, high$-z$ GRB~050904.
These data are coarsely grouped into
three time intervals by \citet{campana06c} and fit to show a time-decreasing X-ray
column density.  At finer time resolution (left plot), we see that the $N_H$ parameter
decreases toward the late time value before and after an unphysical increase.
The maximum $N_H$ corresponds to $E_{\rm peak}$ in the XRT band (right plot).
The hardness during this period tracks the
flux to the $0.6\pm0.2$ power (see also, Figure X in Butler \& Kocevski 2007b), 
consistent with a Band model spectrum evolving via the curvature-effect (Section \ref{sec:interp}).}
\label{fig:050904}
\end{figure}

Several studies have claimed recently a decreasing $N_H$ based on fits
to the XRT data \citep[][GRBs 050730, 050716, and 050904, respectively]{starling05,rol06,campana06c}. 
Each study presents a coarsely time-resolved set of spectral fits, which demonstrate a higher
$N_H$ at early times.  
This is an artificial feature that we observe in fits to most \swift~early afterglows.  It is
especially clear in the brightest afterglows, which often sample the declining tail of the prompt emission.
For each of the 3 bursts with claimed $N_H$ variations (e.g., Figure \ref{fig:050904}), a fine times-scale spectral 
analysis reveals an $N_H$ which both increases and decreases in time (following $\Gamma$ and the flux).
Observed drops in $N_H\gtrsim 10^{21}$ cm$^{-2}$ (or $\gtrsim 10^{22}$ cm$^{-2}$ in the rest-frame)
on timescales of $10-100$s are challenging enough, but drops and increases and
drops again on these timescales are unphysical.

We strongly caution against taking the early $N_H$ values at face value.  
Measurements of $N_H$ at $t\lessim 10^4$s will be artifically high.  
Also, although we cannot rule
variations out in all cases, they are not required by the data and they are also not the simplest interpretation
of the data.
Firm measurements of $N_H$ variability will require finely time-sampled broadband data (e.g., Ultra-violet, X-ray, and $\gamma$-ray data)
to disentangle the effects of the evolving Band model spectrum
from the soft X-ray photoelectric absorption.

For those fitting XRT spectra, we recommend measuring $N_H$ at late times ($t\gtrsim 10^4$s) or
performing joint fits at different time intervals with a single $N_H$ parameter shared between multiple
spectra.  Jointly fit with BAT when possible.  Fine time resolution is essential when testing variable $N_H$; it is not sufficient to
fit exponential times powerlaw models or Band models \citep[e.g.,][]{rol06,campana06c} with coarse time resolution.
The hardness ratio can be utilized to diagnose cases where 
inferred $N_H$ values are likely to vary artificially.

\subsection{Is the Early XRT Emission the Same as Prompt GRB Emission?}
\label{sec:xprompt}

We have shown for seven events that the early X-ray spectra require a fit model which also has been shown
to reliably fit all GRBs \citep[e.g.,][]{preece00,kaneko06,frontera2000,sakamoto03}.  The need for such a model is
also clear from hardness variations \citep[see, also,][]{butNkoc07} and time correlated $N_H$--$\Gamma$ variations observed
for even low signal-to-noise afterglows, which demonstrate a characteristic increase
in spectral curvature in the XRT band.

In cases where $E_{\rm peak}$ is well measured, or using $HR$ when $E_{\rm peak}$ is 
poorly measured, we observe a hard-to-soft evolution and a strong hardness--intensity correlation,
also commonly seen in GRB pulses.
Our correlation can be described as a hardness which tracks the flux to a power $0.43\pm0.07$.
From the Band model fits, our best fit $E_{\rm peak}$--$F$ relation is $0.7\pm0.2$ (Table 3).  
A closely consistent powerlaw relation exists for most GRB pulses, also with a large scatter in observed
values \citep{gol83,karg95}.  The scatter is apparently minimized for bolometric measures
of flux \citep{bnr01}, yielding $E_{\rm peak} \propto F_{\rm bol}^{0.5\pm0.2}$.
The fact that we have observed a consistent relation can be turned around
to imply GRB-like emission with $E_{\rm peak}\approx 1$keV, typically.   That \swift~observes bright X-ray flares appears to
be a consequence of this and also due to the surprising fact that the afterglow is faint at these
times.  It is interesting to speculate that there may be bright optical flares due to internal shocks at times of several
hours after some GRBs with faint afterglows.

The typical $E_{\rm peak}$ values for the XRT are two orders of magnitude below
the mode of the {\it BATSE}~distribution \citep{preece00,kaneko06}.  As we discuss below, some of the
soft $E_{\rm peak}$ values may be due to viewing effects of delayed emission with an
intrinsically higher $E_{\rm peak}$.  However, the soft flare emission implies intrinsic
spectral evolution or soft late central engine activity which would extend the {\it BATSE}~$E_{\rm peak}$
distribution.  Our derived values for $\alpha$ are poorly constrained, but likely consistent with the 
{\it BATSE}~distribution.  Finally, it is remarkable that very soft emission is observed in a
few cases, extending the distribution in $\beta$ to very low values $<-6$ (Figures \ref{fig:4back_spec_panel} and 
\ref{fig:gammas}; GRBs 050714B and 050822 discussed in Paper I).

\subsection{Interpretation of the Spectral Variations}
\label{sec:interp}

Although intrinsic spectral evolution is likely also present, most of the softening trend and
hardness--intensity correlation in GRB pulses is attributed to the so called
``curvature effect'' \citep{fenimore96,snp97,norris02,rnp02,krl03,qin04,ql05,shen05}.
This is also the widely-accepted explanation for the rapid decline X-ray tails of the
prompt emission \citep{nousek06,zhang06,pain07}.   
Derivations from first principles of the
curvature effect on the observed spectra can be found in \citet{gps99,wl99}.

If we imagine a spherical emitting shell at radius $R$ that emits as a delta function at $t_{\circ}$,
the spectral flux $F_E$ scales with the Doppler factor $\delta$ as $F_E \propto F_E[E\delta]/\delta^2$.
Here, $\delta \equiv \gamma (1-\beta_c\cos(\theta))/(1-\beta_c) \approx 1 + \gamma^2\theta^2$, where $\theta$ is 
the viewing angle to emitting material off the line of site.  The photons from larger
angles will be delayed in time $t-t_{\circ} = (1+z)(\delta-1)(1-\beta_c)R/(c\beta_c) \approx (1+z)\theta^2 R/(2c) \propto \delta$.

For a powerlaw spectrum $F_E\propto E^{1-|\alpha|}$, the observed flux declines in time as a powerlaw $(t-t_{\circ})^{-|b|}$,
with $|b| = 1 + |\alpha|$ and no hardness evolution \citep{knp00}.  
For a Band spectrum, we see either the low energy index $\alpha$ or the high energy index $\beta$ or some average of the two,
depending on the location of $E_{\rm peak}$ with respect to the bandpass.  
$E_{\rm peak}$ will decline as $(t-t_{\circ})^{-1}$.
When $E_{\rm peak}$ is in the band,
the $\nu F_{\nu}$ turnover implies $-\alpha_{\rm eff}\approx 1-2$, and we expect to see a powerlaw hardness--intensity 
correlation $E_{\rm peak} \propto F^{0.3-0.5}$.  Larger values of the index are favored observationally, because
they correspond to a higher flux.
We will observe the hardness (which our simulations show to scale linearly with $E_{\rm peak}$
for a range of Band model parameters) to approximately linearly correlate with the fluence.
Departures from this expected behavior will occur
for emitting shells of different shape, for an inhomogeneous emitting surface, for non-instantaneous emission,
or if intrinsic spectral evolution dominates.
Also, the measured flux decay in time is a strong function
of the assumed $t_{\circ}$ \citep[e.g.,][]{liang06}.

\begin{figure}[H]
\centerline{\includegraphics[width=6.0in]{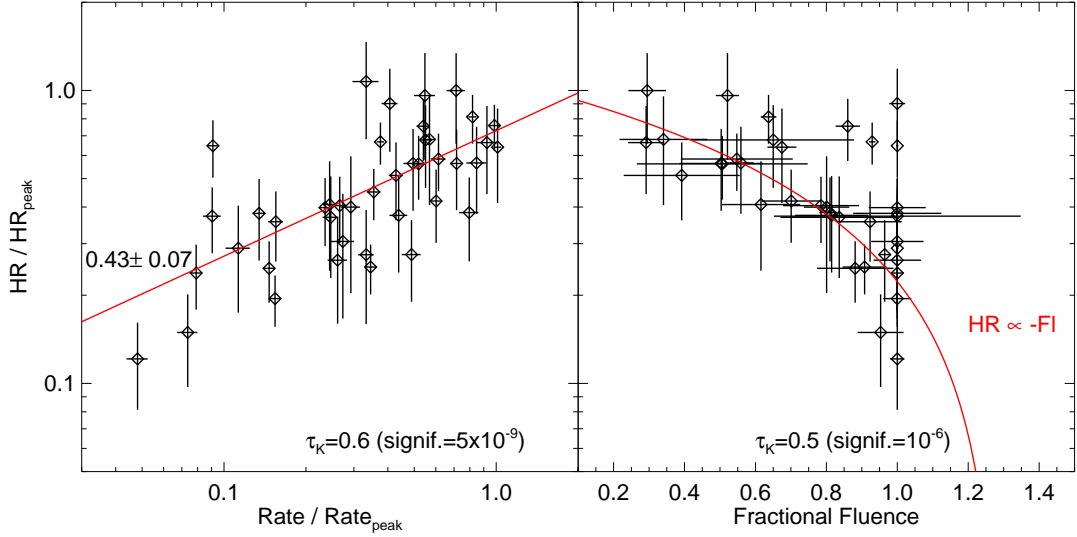}}
\caption{\small
During the decline phase of the X-ray flares from Table 2 (also Figure \ref{fig:nh_vs_gamma}),
the hardness ratio ($HR$), defined as the ratio of counts in the
1.3--10.0 keV band to the counts in the 0.3--1.3 keV band,
correlates strongly (Kendall's $\tau_K=0.6$)
with the count rate (0.3--10.0 keV), following
roughly a powerlaw relationship (left plot).  There is a consistent and long known relation valid
for a majority of pulses seen in GRBs \citep{gol83,karg95,ford95,bnr01}.
The hardness also correlates strongly with the fluence (right plot), as is also the
case for GRBs \citep{lnk96,ryde05}.  That is, the hardness evolves more rapidly
when the flares are brighter.}
\label{fig:hard_flares}
\end{figure}

Our best fit $HR$--$F$ relation index (Figure \ref{fig:hard_flares}) and
our average $E_{\rm peak}$--$F$ relation relation index (Section \ref{sec:xprompt}; Table 3)
are consistent with those expected in this simple picture.  Spectral variations
are not inconsistent with the curvature effect, as recently suggested by \citet{zlz06}.  Rather, they facilitate a higher order test
of the curvature effect, and allow us to confirm the curvature effect in way
that shows the X-ray phenomenology to closely parallel the $\gamma$-ray phenomenology.  
Moreover, the scatter in our $HR$--$F$ relation
(Figure \ref{fig:hard_flares} left) is less than that found for time-index--energy index relations \citep{nousek06,pain07}, which assume
powerlaw X-ray spectra.  

The mean time index for the $E_{\rm peak}$ decays in Table 3 is $-1.4\pm0.6$, consistent with unity.  This indicates that
our choice to associate $t_{\circ}$ with the start of the flare or pulse is roughly correct, in agreement with the findings of \citet{liang06}.
Although we see evidence that later flares often have lower $E_{\rm peak}$ in the same event with multiple flares (e.g., Figure
\ref{fig:3main_spec_panel}),
we do not see a correlation between $t_{\circ}$ in Table 3 and $E_{\rm peak}$ just after that time.

We have observed two cases of $\alpha$ evolution (Figures \ref{fig:3main_spec_panel} right) which 
accompanies the $E_{\rm peak}$ evolution.  Due to the proximity of $E_{\rm peak}$ to the bottom of the XRT pass
band and also due to the possibility of a modestly incorrectly measured $N_H$, 
these cases should be interpreted cautiously.
This evolution, or that observed for $\beta$ (Section \ref{sec:xprompt}) 
cannot be accounted for by the curvature effect and must be intrinsic.

In most cases, the X-ray light curve is simply declining early on (possibly with weak flaring superimposed),
and we observe approximately secularly declining $E_{\rm peak}$ and $HR$ values.  In a handful of cases where multiple flares
follow a GRB (e.g., Figure \ref{fig:main_hard_panel}, 060124; and \ref{fig:060714_both}, 060714), the hardness tracks the flux both upward and
downward.  Because the brightest case (060124 in Figure \ref{fig:3main_spec_panel} right) also shows upward and downward $E_{\rm peak}$ trends, we
believe this behavior is likely responsible for the $HR$ evolution.  The parallel or overlapping tracks observed here for bursts with multiple flares on the
$HR$--$F$ diagram is also seen for GRB pulses \citep{bnr01}.

\section{Conclusions}
\label{sec:concl}

We have measured the spectral evolution properties for GRBs and afterglows in
the Swift sample, taken prior to and including GRB~061210.  
We have established similar spectral evolution properties between the X-ray emission 
coincident with two GRBs (060124 and 061121) and the X-ray emission in the rapid declines
following several GRBs and in 27 flares ocurring $10^2-10^3$ after their GRBs.

Indirectly from absorbed powerlaw fits
which show a time-variable $N_H$ and directly from Band model fits, we have derived
constraints on the $\nu F_{\nu}$ spectrum peak energy $E_{\rm peak}$.  We observe this
quantity to evolve in time and to typically cross the XRT bandpass during the
early X-ray afterglow.  Because the X-ray hardness changes little for Band spectra with
$E_{\rm peak}$ outside the bandpass, the strong hardness variation we observe in $>$90\%
of \swift~early afterglows \citep{butNkoc07} imply $E_{\rm peak}\approx 1$ keV, typically.
We observe this evolution in data taken in both the WT and PC modes (e.g., 050607 and 050714B)
and following both long duration and short duration (e.g., 050724 and 051227) GRBs.
The hardness ratio and $E_{\rm peak}$ values scale with the flux as would be expected
from the relativistic viewing effects of an expanding fireball.  
This implies that the true variability timescale is {\it even shorter}~than that 
measured from the observed flare durations.

Because the late flares
are typically softer than the GRB emission, and because the Band model $\alpha$ and
$\beta$ parameters also appear to evolve in some cases, there is likely an intrinsic
evolution of the fireball.  If the flares are due to shells moving out with lower
bulk Lorentz factor or at larger radii than for the prompt emission, we may expect to
see differences in the time properties of flares observed at different epochs.  This will
be explored in a separate paper \citep{koc07}.  If the evolution is occurring on longer
timescale at later times, when the sensitive XRT is observing, the early X-ray afterglows
would provide a unique test-bed for theories explaining GRBs, the emission mechanisms, and
possibly the progenitors.
The internal shocks must be active after $10^3$s and must be able to produce emission with
$E_{\rm peak}\approx 1$ keV and very soft $\beta\lessim -6$ \citep[see, also,][]{zhang06}.  
Especially relevant to the Gamma-ray Large Area Telescope (GLAST),
electrons energized by the X-ray flares may Compton upscatter photons at larger radii or in
the external shock to the $\gamma$-rays \citep{kocetal07}.

\acknowledgments
N.~R.~B gratefully acknowledges support from a Townes Fellowship at U.~C. Berkeley Space Sciences Laboratory and partial support
from J. Bloom and A. Filippenko.  
D.~K. acknowledges financial support through the NSF Astronomy $\&$ Astrophysics Postdoctoral Fellowships under award AST-0502502.
This work 
 was conducted under the auspices of a DOE SciDAC grant (DE-FC02-06ER41453), which provides support to J. Bloom's group.
Special thanks to the {\it Swift}~team for impressively rapid public release and analysis of the XRT data.  
Thanks to J. Bloom and the U.~C. Berkeley GRB team for comments on the manuscript and several useful
conversations.  We thank an anonymous referee for a very useful and critical reading of the manuscript.

\begin{deluxetable}{cccccccc}
\tabletypesize{\small}
\tablewidth{0pt}
\tablecaption{Selected Band or Powerlaw*Exponential Model Spectral Fits}
\startdata\hline\hline
GRB 	& Time 	&  $\alpha$   	 & $\beta$       & $E_{\rm peak}$      & 0.3-10 keV Flux 			& $\chi^2_{\nu}$ ($\nu$) 	& Signif. \\
        & [s] 	&              	 &               & [keV]                         & [$10^{-9}$ erg cm$^{-2}$ s$^{-1}$]   &		&  \\\hline

060124 & 569--600 & $-1.23\pm0.04$ & ... & $108^{+\infty}_{-22}$ & $26.8\pm1.0$ & 1.21 (154) & $5.9\sigma$ \\
060124 & 400--569 & $-1.04\pm0.03$ & $-2.0^{+0.0}_{-0.1}$ & $27.2^{+3.3}_{-1.6}$ & $8.6^{+0.2}_{-0.1}$ & 1.01 (410) & $10^{-77}$ \\
060124 & 720--770 & $-0.3^{+0.3}_{-0.6}$ & $-2.2\pm0.1$ & $1.3\pm0.2$ & $5.8^{+0.3}_{-0.2}$ & 1.08 (158) & $10^{-12}$ \\\hline

061121 &  60--90  & $-1.12^{+0.01}_{-0.02}$ & ... & $270^{+\infty}_{-40}$ & $55.3^{+1.0}_{-1.3}$ & 1.07 (270) & $9.3\sigma$ \\
061121 & 126--140 & $-0.0^{+0.0}_{-0.9}$ & $-2.4^{+0.1}_{-0.2}$ & $0.95^{+0.05}_{-1.0}$ & $3.7\pm0.2$ & 1.16 (117) & $3.9\sigma$ \\\hline

060614 &  97--111 & $-0.7\pm0.1$ & $-2.4\pm0.1$ & $8.6\pm1.2$ & $59^{+3}_{-2}$ & 0.86 (169) & $10^{-98}$ \\
060614 & 237--297 & $-1.2\pm0.2$ & $-2.8^{+0.2}_{-0.3}$ & $1.1\pm0.1$ & $3.2\pm0.1$ & 0.97 (246) & $10^{-27}$ \\\hline

\enddata

{Note.---The quoted errors correspond to the 90\% confidence region.  The ``Signif.'' column
refers to the fit improvement significance relative to a simple powerlaw model,
determined from a $\Delta \chi^2$ test.  The quoted fluxes are unabsorbed.}
\label{table:spec_pars}
\end{deluxetable}

\begin{table}[H]
\begin{center}
\caption{27 Bright XRT Flares}
\vspace{5mm}
\footnotesize
\begin{tabular}{lclc}\hline\hline
 GRB & Time Region [s] & GRB & Time Region [s]\\\hline
050502B  & 400--1200 & 050712 & 150--300 \\
050730 & 130--300 & 050730 & 300--600 \\
050730 & 600--800 & 050822 & 410--650 \\
050904 & 350--600 & 051117A & 1250--1725 \\
051117A & 800--1250 & 060111A & 200--500 \\
060124 & 300--650 & 060124 & 650--900 \\
060204B & 100--270 & 060204B & 270--450 \\
060210 & 100--165 & 060210 & 165--300 \\
060210 & 350--450 & 060418 & 83--110 \\
060607A & 93--130 & 060607A & 220--400 \\
060714 & 100--125 & 060714 & 125--160 \\
060714 & 160--230 & 060729 & 156--300 \\
060904A & 250--600 & 060904A & 600--1000 \\
060904B & 140--300 & & \\\hline
\end{tabular}
\end{center}
\label{tab:flares}
\end{table}

\begin{table}[H]
\begin{center}
\caption{$E_{\rm peak}$ Evolution Properties}
\vspace{5mm}
\footnotesize
\begin{tabular}{lcccc}\hline\hline
 GRB & $t_{\circ}$ [s] & Time Index & Flux Index & Data Points Fit \\\hline
060124	& 510	&		$-2.9\pm1.3$ &	$-3.6\pm1.7$ & 2 \\
060124	& 555	&		$-2.2\pm0.3$ &	$-1.8\pm0.5$ & 3 \\
060124	& 567	&		$-0.8\pm0.1$ &	$-0.3\pm0.1$ & 4 \\
060124	& 685	&		$-2.2\pm0.4$ &	$-1.2\pm0.2$ & 3 \\
060526	& 240	&		$-1.2\pm0.1$ &	$-1.0\pm0.1$ & 5 \\
060614	& 0	&		$-2.1\pm0.1$ &	$-0.72\pm0.03$ & 19 \\
060729  &  75 	&		$-2.0\pm0.5$ &  $   -0.4\pm0.1$ & 5 \\
060729	& 155	&		$-0.7\pm0.2$ &	$-0.4\pm0.1$ & 4 \\
060904B	& 140	&		$-1.3\pm0.3$ &	$-0.7\pm0.2$ & 5 \\
060929	& 470	&		$-1.1\pm0.2$ &	$-0.6\pm0.1$ & 7 \\\hline
\end{tabular}
\end{center}
{\footnotesize Notes: Changes in the best-fit $E_{\rm peak}$ with time are relative to the start $t_{\circ}$.
The start time is somewhat arbitrary, based on the approximate start of each pulse (or flare).}
\label{tab:epvals}
\end{table}

\end{document}